\input{aipcheck}

\documentclass[
    ,final            
  ]
  {aipproc}

\layoutstyle{8x11double}

\begin{document}

\title{High energy emission from massive stars: the precocious  X-Ray recovery of  Eta Carinae after January 2009 minimum}


\author{Elena Pian}{
address={INAF, Trieste Astronomical Observatory, Via G.B. Tiepolo, 11, I-34143 Trieste, Italy}
}

\author{Sergio Campana}{
address={INAF, Brera Astronomical Observatory, Via E. Bianchi, 46, I-23807, Merate (LC), Italy}
}

\author{Guido Chincarini}{
address={INAF, Brera Astronomical Observatory, Via E. Bianchi, 46, I-23807, Merate (LC), Italy}
}

\author{Michael F. Corcoran}{
address={NASA Goddard Space Flight Center, Greenbelt, MD 20771, USA}
}

\author{Kenji Hamaguchi}{
address={NASA Goddard Space Flight Center, Greenbelt, MD 20771, USA}
}

\author{Theodore Gull}{
address={NASA Goddard Space Flight Center, Greenbelt, MD 20771, USA}
}

\author{Paolo A. Mazzali}{
address={INAF, Padova Astronomical Observatory, Vicolo dell'Osservatorio, 5, I-35122 Padova, Italy}
}

\author{Christina C. Thoene}{
address={INAF, Brera Astronomical Observatory, Via E. Bianchi, 46, I-23807, Merate (LC), Italy}
}

\author{David Morris}{
address={NASA Goddard Space Flight Center, Greenbelt, MD 20771, USA}
}

\author{Neil Gehrels}{
address={NASA Goddard Space Flight Center, Greenbelt, MD 20771, USA}
}

\begin{abstract}
 
We observed the massive binary stellar system of Eta Carinae in the 0.3-10 keV energy range with the X-ray Telescope onboard the Swift satellite during the period 15 December 2008 - 11 March 2009, i.e. 1 month before to 2 months after the  X-ray drop from maximum to minimum, thought to be associated with the periastron encounter of the primary star by the hot companion.  Beginning a few months before eclipse, the interaction between the winds of the two stars intensifies and the X-ray flux reaches maximum.  The flux drops dramatically thereafter, subsiding in about 20 days to a level that is at least a factor 10 lower than the 'high state', i.e. the X-ray emission state of the system during the largest fraction of its  5.52 yr orbit ($\sim 10^{-11}$ erg~s$^{-1}$~cm$^{-2}$).    Unlike in previous cycles, when the low state lasted about 2.5 months, 
observations with RXTE showed that the X-ray flux started its recovery
to  normal level about 1.5 months after the minimum.   We suggest that this early recovery may be due to the fact that the companion wind reaches terminal velocity before encountering the shock. 

 \end{abstract}

\maketitle


\section{Introduction}

Eta Carinae  (estimated distance: 2.3 kpc) is a violently unstable, extremely luminous object ($5 \times 10^6$ L$_\odot$), and very massive, i.e. the typical potential progenitor of a powerful supernova and possibly of a Gamma-Ray Burst or X-ray Flash.  The system is believed to have had an initial mass larger than 150 M$_\odot$ (Davidson \& Humphreys 1997; Hillier et al. 2001).  It is currently in a short, poorly understood evolutionary stage, known as the Luminous Blue Variable phase, which is thought to occur near the onset of pulsational instabilities.  In this transient state, it provides a convenient laboratory to study how extremely luminous, massive stars evolve and how they shape their environments both geometrically and chemically.

Eta Carinae is best known for an extraordinarily powerful eruption in 1843 that sent  $>12$ M$_\odot$  of its carbon and oxygen-depleted atmosphere into space (Smith et al. 2003a), creating the Homunculus nebula.  Eta Carinae also had a minor eruption ($\sim1 M_\odot$) in 1890, which produced a small bipolar nebula inside the Homunculus (the 'Little Homunculus'; Ishibashi et al. 2003).  The star still exhibits a strong mass loss ($\sim10^{-3}$ M$_\odot$ yr$^{-1}$) preferentially in the polar direction (Smith et al. 2003b).   

The strength of some narrow lines of Eta Carinae, notably HeI $\lambda$10830, vary predictably with a period of 5.52 yr (Damineli 1996).  Observations at optical, infrared, radio and X-ray frequencies have shown a periodic behavior as well.     ROSAT observations in 1992 first showed a variation that appeared to be correlated with Damineli's emission-line variations (Corcoran et al. 1995), and subsequently the 2-10 keV light curve of the star obtained by the RXTE showed in detail the X-ray variation culminating in a rapid, unstable rise to maximum and steep fall to minimum lasting for $\sim$3 months, a shorter time interval compared to the optical spectroscopic minimum (Corcoran 2005; Damineli et al. 2008).  Eta Carinae has been detected and resolved from nearby sources also in hard X-rays (20-100 keV) by the IBIS/ISGRI camera onboard INTEGRAL (Leyder et al. 2008), and possibly detected above 100 MeV by AGILE (Tavani et al. 2009).

Based on previous monitoring in X-rays with RXTE (Corcoran 2005) and at lower frequencies, an X-ray minimum was expected to occur around  mid January 2009.  Therefore, we  organized X-ray observations with various satellites including Swift.  In Section 2 we describe the Swift X-ray Telescope (XRT) observations and in Section 3 we report and discuss the most important  results.

\section{Swift/XRT observations and data analysis}

We started observing Eta Carinae with the Swift satellite (Gehrels et al. 2004)  in mid December 2008 with the XRT (Burrows et al. 2005) with a cadence of about 3-4 days, until 14 January 2009, when the X-ray flux - as monitored with the RXTE satellite - became too faint to be studied with the same accuracy as in previous observations.
We resumed XRT observations on 11 March 2009, when  RXTE detected early recovery from the deep minimum state, expected to occur  one month later.   Thereafter, we have continued observing Eta Carinae with XRT at a monthly cadence.   An exposure time of 4 ks was adopted for all observations, except on 14 January 2009, when, because of its anticipated low flux level, the source was observed for 15 ks.  Observations with the Ultra-Violet Optical Telescope (Roming et al. 2005) were not possible because of the excessive brightness of the target.   We report here the preliminary outcome of the XRT observations accomplished up to March 11, 2009.  Results of the more recent XRT observations and of the Burst Alert Telescope (Barthelmy et al. 2005) observations will be reported elsewhere.     

The Window Timing mode for XRT observations was selected to avoid optical loading.  The data were processed  with standard procedures (see e.g. Campana et al. 2008), filtering and screening criteria using the FTOOLS in the Heasoft package (v. 6.3.1).   Spectra were extracted for each observation; we used the latest spectral redistribution matrices in the Calibration Database maintained by HEASARC.  All spectra were rebinned with a minimum of 20-30 counts per energy bin (depending on
the source strength with respect to the background) to allow $\chi^2$ fitting with XSPEC.  
The spectra were  fitted with two "mekal" components, i.e. thermal emission from plasma emitting in two regions with different temperatures: a temperature of $\sim$0.5 keV accounts well for the softer part of the spectrum (up to 2 keV), while above 2 keV the temperature is $\sim$5 keV, in agreement with previous spectroscopic studies of the source  with the  XMM-Newton EPIC instrument (Hamaguchi et al. 2007).     

The XRT spectra  are presented, with their fitting curves, in Figure 1.   The fit parameters of the 14 January 2009 spectrum (start of the deep minimum) are very poorly constrained because of the low signal,  dominated here by the constant background of the Eta Carinae nebula.   
The softer component is modified by absorption with a constant equivalent hydrogen column density of $1.2 \times 10^{22}$ cm$^{-2}$, while the equivalent hydrogen column density of the harder component varies significantly. It exhibits a dramatic variation between January and March, straddling the minimum:  in the 11 March 2009 spectrum the $N_H$ fitted value of the harder component is $\sim 10^{23}$ cm$^{-2}$, i.e. nearly a factor 10 larger than prior to minimum.



\begin{figure}
 \includegraphics[angle=90,angle=90,angle=90,height=.6\textheight]{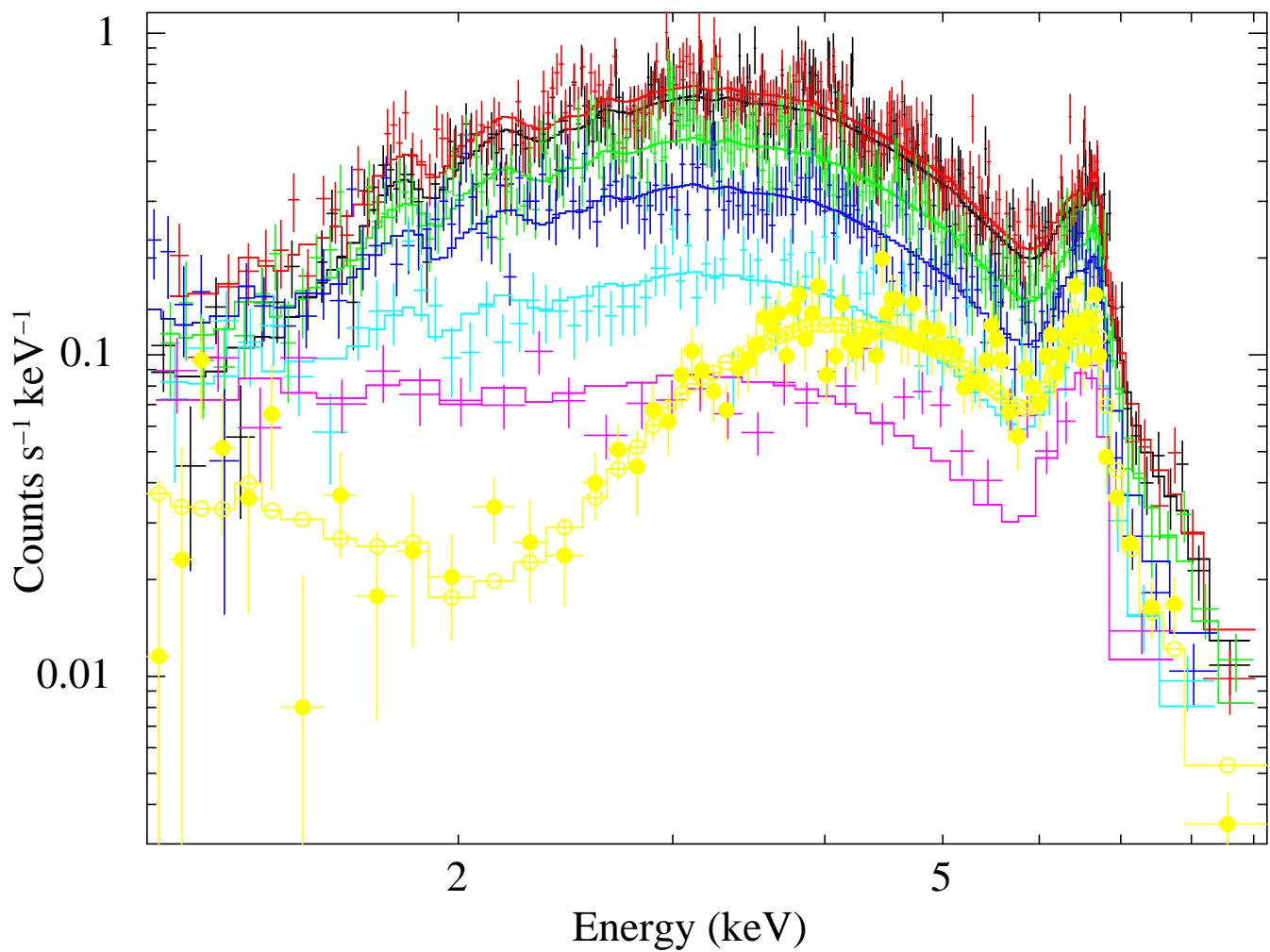}
 \caption{Swift/XRT spectra of Eta Carinae acquired from Dec 2008 to Mar 2009, covering the phase preceding the minimum and the recovery after minimum of X-ray emission: 2008 Dec 19 (black), Dec 23 (red), Dec 27 (green), Dec 31 (blue), 2009 Jan 04 (cyan), Jan 07 (purple), March 11 (yellow circles).  The last epoch spectrum was taken after the deep minimum occurred on Jan 11, and indicated a faster rate of X-ray flux increase than expected based on the previous monitored X-ray cycles of Eta Carinae. Note the heavy absorption in the March 11 spectrum: the value of the equivalent hydrogen column density  ($\sim 2 \times$10$^{23}$ cm$^{-2}$)   is almost an order of magnitude larger than before the minimum, a behavior already known from Chandra and XMM-Newton observations from the past cycle (Hamaguchi et al. 2007).   Note also the fluorescent Fe line. 
 }
\end{figure}



\section{Results and discussion}

The unabsorbed flux of Eta Carinae measured by XRT in the 0.3-10 keV range before the 14 January 2009 is about $10^{-10}$ erg~s$^{-1}$~cm$^{-2}$. It drops rapidly by more than one order of magnitude in the last 20 days before the minimum.  
While the flux subsided as predicted, it returned to the normal 'high-state' level at least one month earlier than expected based on the previous cycles that had been monitored by the RXTE (\url{http://asd.gsfc.nasa.gov/Michael.Corcoran/eta_car/etacar_rxte_lightcurve/index.html}).

The multiwavelength monitoring of  Eta Carinae  indicates a fundamental flux periodicity of 5.52 yr and strongly suggests that it is a binary system. In a binary model, most of the dramatic multiwavelength changes are believed to be produced by the interaction of the faster wind and the far-ultraviolet flux from a hot companion star (see Iping et al. 2005) in a highly eccentric orbit  (Pittard et al. 1998; Gull et al. 2009, and references therein). Variations in the X-rays are produced by  the  collision of the slower, more massive wind of Eta Carinae with the faster, less massive wind of the companion star. A current guess at the system parameters describes a massive hot companion with M $\sim$ 30 M$_\odot$ plus a brighter primary star with M  $\sim$ 80 M$_\odot$ in a highly eccentric (e > 0.9), 5.52 yr orbit.    The X-ray minima and correlated low optical spectroscopic states are thought to arise during periastron passage when the wind-wind interaction region collapses and the far-ultraviolet radiation is absorbed by the primary wind.  The overall X-ray brightness variations are explained well by the colliding wind mechanism (Parkin \& Pittard 2008; Parkin et al. 2009).

However, the dramatic changes in X-ray flux as the emission increases to maximum, the variations of the absorbing material in front of the X-ray emitting region near the X-ray minimum, the nature of the rapid fall from X-ray maximum to minimum, and the excess in equivalent hydrogen column density after recovery are not well understood.

The Swift observation  of 11 March 2009 is  consistent with the transition from 'deep' to 'shallow' minimum seen in the XMM-Newton spectra during the 2003 campaign (Hamaguchi et al. 2007).  The latter indicated
that during 'deep minimum'   the wind-wind shock is basically disrupted and that during the 'shallow minimum' transition the shock re-forms.  Thus, the observed early 'recovery' suggests that the companion wind shock has re-established earlier than in previous cycles.  Hamaguchi et al. (2007) suggested that the ultraviolet flux from Eta Carinae could decelerate the companion's wind near periastron and weaken the shock. When the stars are close, only a portion of the companion's wind reaches terminal velocity before it encounters the shock, because of radiative braking.  If the entire companion wind reaches terminal velocity before encountering the shock, higher emission is produced at an earlier epoch during the cycle, similar to that seen during the 'full recovery'.
In view of the strength that the high energy component of Eta Carinae may attain, this source is a primary target for multiwavelength monitoring.


\begin{theacknowledgments}
EP, SC, PM and CT would like to thank the organizers for a stimulating and successful meeting.   We are grateful to the Swift team for excellent support of these observations.  Financial support from INAF through PRIN 2006 and ASI  through contract I/088/06/0 is acknowledged.
\end{theacknowledgments}



\bibliographystyle{aipproc}   
\vspace{-0.3cm}

\bibliography{sample}

\IfFileExists{\jobname.bbl}{}
 {\typeout{}
  \typeout{******************************************}
  \typeout{** Please run "bibtex \jobname" to optain}
  \typeout{** the bibliography and then re-run LaTeX}
  \typeout{** twice to fix the references!}
  \typeout{******************************************}
  \typeout{}
 }

\end{document}